\documentclass[sigconf]{acmart}

\usepackage{subfig}

\def\BibTeX{{\rm B\kern-.05em{\sc i\kern-.025em b}\kern-.08emT\kern-.1667em\lower.7ex\hbox{E}\kern-.125emX}}
    
\setcopyright{none}
\renewcommand\footnotetextcopyrightpermission[1]{} 
\begin{document}

%
\title{Introducing Semantic Capability in LinkedIn's Content Search Engine}

%
\author{Xin Yang, Chujie Zheng, Madhumitha Mohan, Sonali Bhadra, Pansul Bhatt, Lingyu (Claire) Zhang, Rupesh Gupta}
\affiliation{%
  \institution{LinkedIn Corporation}
  \city{Mountain View, CA}
  \country{USA}}
\authornote{Corresponding author: rugupta@linkedin.com}


%
\renewcommand{\shortauthors}{Gupta, et al.}

%
\begin{abstract}
In the past, most search queries issued to a search engine were short and simple. A keyword based search engine was able to answer such queries quite well. However, members are now developing the habit of issuing long and complex natural language queries. Answering such queries requires evolution of a search engine to have semantic capability. In this paper we present the design of LinkedIn's new content search engine with semantic capability \footnote{https://www.linkedin.com/search/results/content}, and its impact on metrics.
\end{abstract}

\maketitle

\section{Introduction}
\label{sec:introduction}
LinkedIn operates the world's largest professional networking platform, and provides members the ability to create and consume professional content through various channels like our Feed, Notifications and Search. Over the last few years, we have seen an increase in the use of Search for consumption of content. This has been accompanied by an increase in the complexity of search queries, in terms of their length, the usage of natural language (e.g. "how to ask for a raise?"), as well as included concepts (e.g. "dropout in AI").

Taking a look at our own capabilities, we observed that we had some room for improving our content search results for complex queries. At times, we were either returning no posts since we did not have any posts that contained all of the keywords in a query, or we were returning posts that contained all of the keywords in the query but did not correctly answer the question due to our lack of conceptual understanding of the query.

However, our analysis showed that we often did have posts in our search index that could provide a correct answer, even if they didn't contain all of the keywords in the query. This motivated us to introduce semantic matching capability in our content search engine. In this paper, we describe the design of this new content search engine.

\section{Objectives}
\label{sec:objectives}

The objective of our content search engine is to serve high-quality engaging posts for every query. We measure our progress towards this objective through two quantifiable metrics:

\begin{enumerate}
    \item On-topic rate (a quality metric): We ask a GPT to assign each returned post a label of 1 if it is well written and answers the query, and 0 otherwise. Then, the on-topic rate metric is computed as the percentage of posts with label 1 among the top-10 posts. 
    \item Long-dwells (an engagement metric): We look at the amount of time spent by the searcher on each returned post. If the time spent is more than N seconds (where N depends on the type of the post) then we assign it a label of 1, and 0 otherwise. Then, the long-dwells metric is computed as the number of posts with label 1. We chose long-dwell to capture engagement rather than an interaction (e.g. liking or commenting) as we observed that searchers hesitate to interact with posts that are created by people they don't know or posts that are old. 
\end{enumerate}

So, in terms of quantifiable metrics, there are two objectives: maximization of on-topic rate and maximization of long-dwells. We designed our content search engine to support optimization of these two objectives.

\section{High-level design}
\label{sec:highlevel}
\begin{figure*}[h]
	\includegraphics[width=0.9\textwidth] {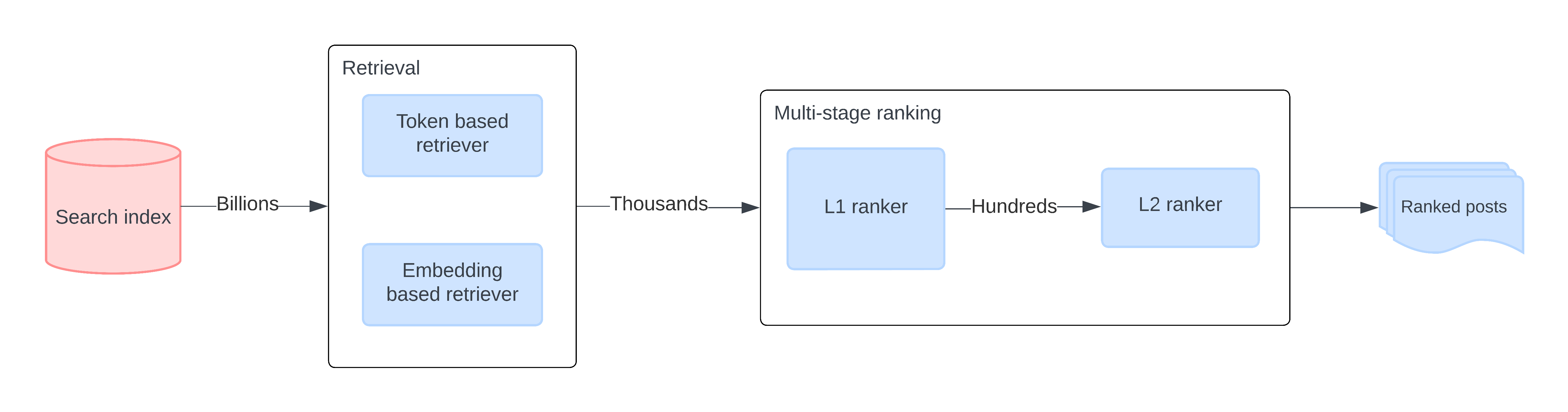}
	\caption[]{High-level design of the content search engine consisting of a retrieval layer and a multi-stage ranking layer.}
	\label{fig:highlevel}
\end{figure*}

Our content search engine has two layers (as shown in Figure~\ref{fig:highlevel}): a retrieval layer and a multi-stage ranking layer. When a query is received, the retrieval layer first selects a few thousand candidate posts from the overall pool of billions of posts. Then, the multi-stage ranking layer scores these candidate posts in two stages and returns a ranked list of posts.

Let's take a look at the design of these two layers in detail.

\section{Retrieval layer}
\label{sec:retrieval}
As shown in Figure~\ref{fig:highlevel}, the retrieval layer consists of two retrievers, a token based retriever (TBR) and an embedding based retriever (EBR).

TBR selects candidate posts that contain the exact same keywords as in the query. In order to do this, we maintain an inverted index that is essentially a mapping from each keyword to a list of posts that contain that keyword. Now when a query is received, then for each keyword in the query we select the list of posts that contain that keyword from the inverted index. Then, we perform an intersection of all these lists to retain only those posts that contain all the keywords in the query.

\begin{figure*}[h]
	\centering
	\includegraphics [width=0.68\textwidth]{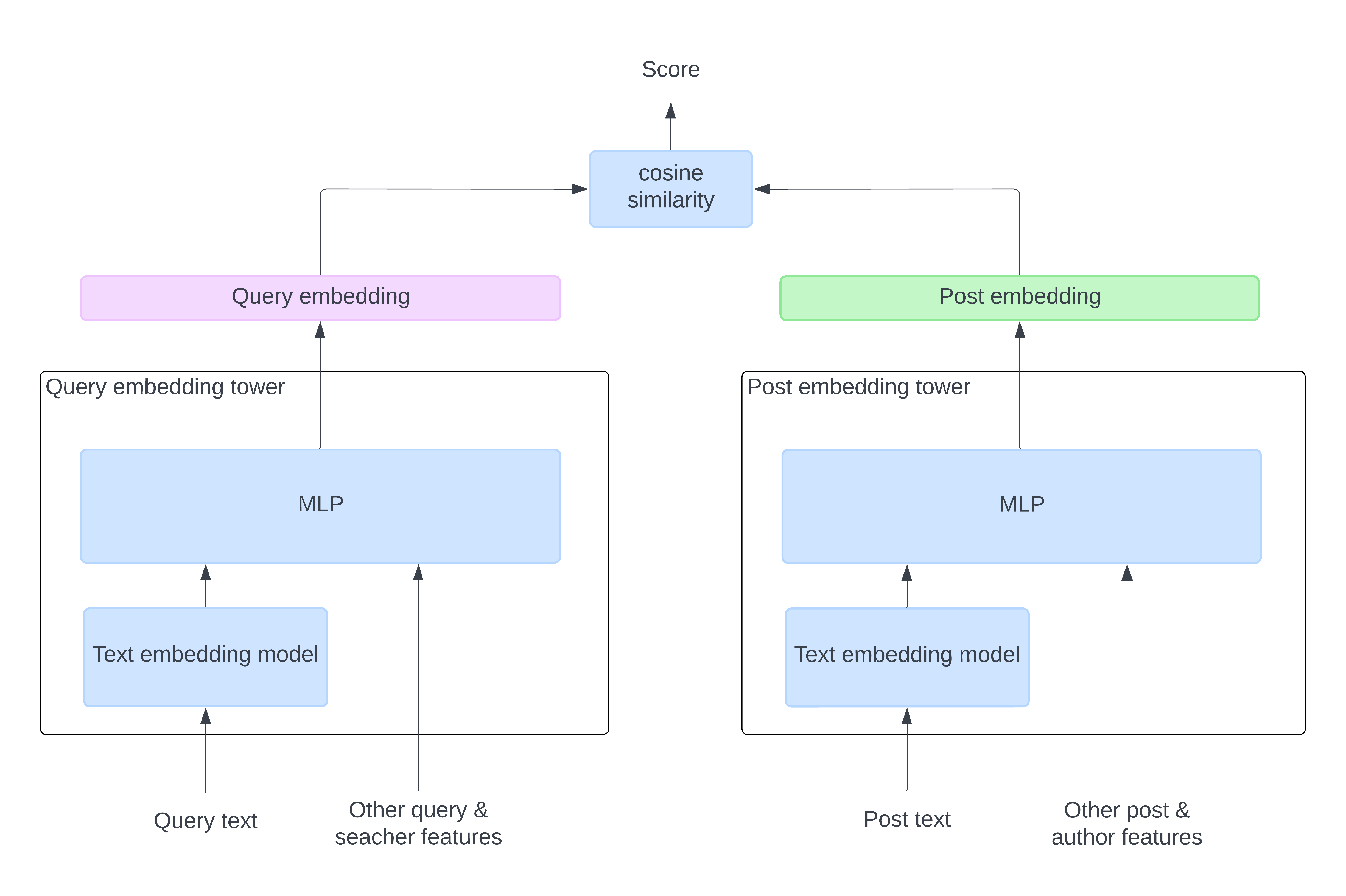}
	\caption[]{Architecture of the two-tower model used in EBR.}
	\label{fig:twotower}
\end{figure*}

EBR, on the other hand, uses an AI model to select candidate posts. The architecture of this model (known as a two-tower model) is shown in Figure~\ref{fig:twotower}. The model consists of two trainable towers: a query embedding tower and a post embedding tower. The query embedding tower takes query text, and a few other features of the query and the searcher as inputs. The query text is passed to a text embedding model (multilingual-e5), which produces an embedding of the query text. This query text embedding is concatenated with the other features and passed to a multi-layer perceptron (MLP) which produces another embedding that we call "query embedding." Similarly, the post embedding tower takes post text, and a few other features of a post and the author as inputs to produce a "post embedding." Then, the cosine similarity between the query embedding and the post embedding is used as the score for the post.

We chose multilingual-e5 \cite{e5} for embedding text as it's an open-sourced model that supports multiple languages and scores well on the MTEB leaderboard \cite{mteb} for the retrieval task.

We train the two towers in this model using (query, post, label) data collected from historical serving of posts for queries issued to the content search engine. We prepare the label as a combination of the on-topicness and long-dwell labels (label aggregation method).

This two-tower model has two beneficial properties:
\begin{enumerate}
    \item For a given query, scoring each post using the model and selecting the top-k posts is the same as finding the k posts whose post embeddings are closest to the query embedding. 
    \item Since there is no interaction between the query and post embedding towers, the post embeddings can be pre-computed and stored for each post.
\end{enumerate}

\begin{figure*}[h]
	\centering
    \includegraphics[width=0.7\textwidth] {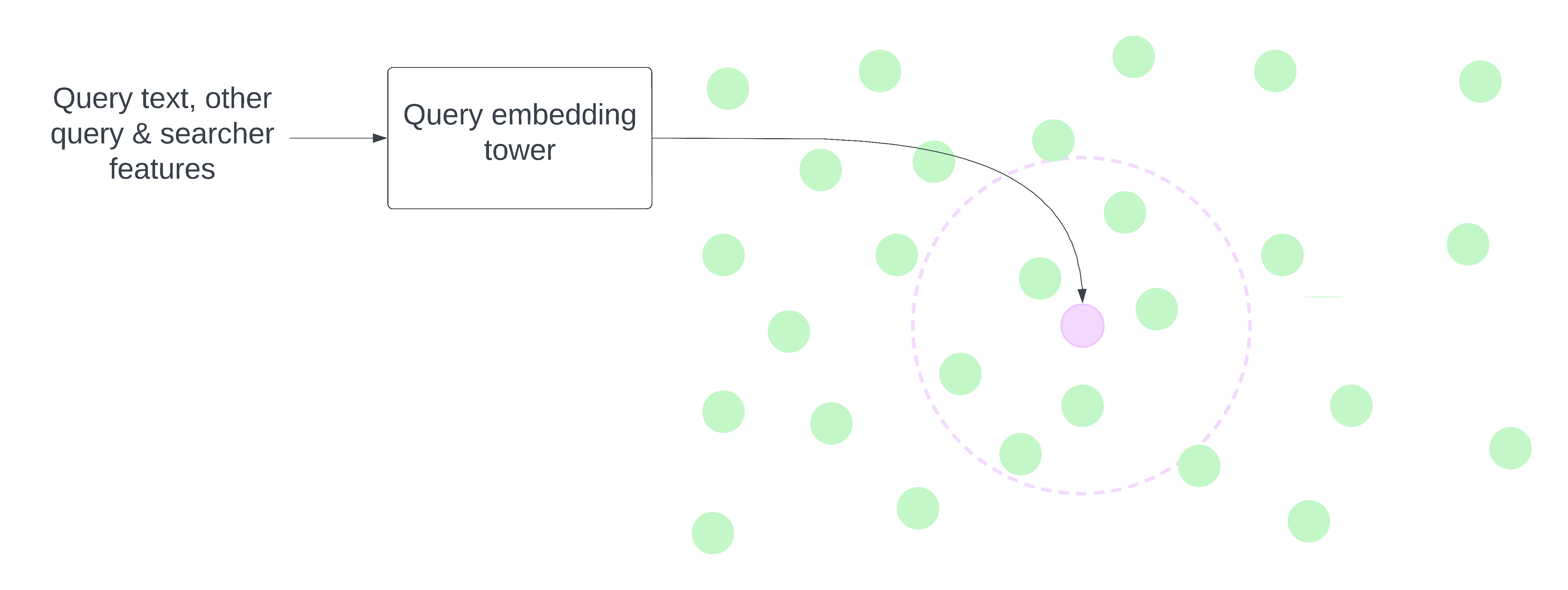}
	\caption[]{Approximate nearest neighbor search in EBR using precomputed post embeddings (green) and real-time computed query embedding (pink).}
	\label{fig:ebr}
\end{figure*}

Due to these two properties we don't have to compute the score of every single post using the model in real-time when a query is received, which would be prohibitively expensive given the billions of posts on LinkedIn. Instead, we pre-compute the post embeddings of all the posts using the post embedding tower and store them in an embeddings store. For this, we run an offline batch job to compute the post embeddings of all the existing posts and push them to the embeddings store. Then, we run a nearline job on Samza \cite{samza} to compute the post embeddings of newly created posts and push them to the same embeddings store. Now, when a query is received, we compute only the query embedding in real-time using the query embedding tower and then use an approximate nearest neighbor search method to select k posts whose post embeddings are closest to the query embedding as shown in Figure~\ref{fig:ebr}. This enables EBR to return candidate posts very efficiently.

EBR has the following advantages compared to TBR:
\begin{enumerate}
    \item It enables semantic matching between the query and posts. This is due to the usage of text embeddings in the model that represent query text and post text as concepts.
    \item It enables personalized selection of posts depending on the searcher. This is because searcher features can be added as inputs to the query embedding tower, which allows selection of different posts for the same query issued by different searchers.
    \item It enables optimization of any set of objectives, as long as the objectives can be incorporated into the label in training data.
\end{enumerate}

However, TBR is still necessary for queries that require exact keyword-based matching, such as a navigational query where a searcher is trying to find a specific post (e.g. "Introducing Semantic Capability in LinkedIn's Content Search Engine"). So, we select a few thousand candidate posts from both EBR and TBR, and pass them to the multi-stage ranking layer.

\section{Multi-stage ranking layer}
\label{sec:ranking}
Since we have fewer posts in the ranking layer, it is possible to score each one in real-time. This means that unlike EBR, the model used in the ranking layer can allow interactions between the query and post features to optimize for on-topic rate and long-dwells. Optimizing for these two objectives is a complex problem. On-topic rate depends on query-post matching and post quality. Long-dwells can depend on those factors as well as searcher intent, searcher-author familiarity, author reputation, post popularity, post freshness and more. Therefore, a complex model is necessary to yield optimal results.

Scoring each of the posts using a complex model can be quite slow. So, we perform ranking in two stages, as shown in Figure 1. In the first stage (also called the L1 ranking stage), we use a simple model to score all the few thousand candidate posts from the retrieval layer and pass the top few hundred candidate posts to the second stage (also called the L2 ranking stage). In the second stage, we use a complex model to score these few hundred candidate posts and order them to prepare the final search results.

\begin{figure*}[h]
	\centering
	\includegraphics[width=0.8\textwidth] {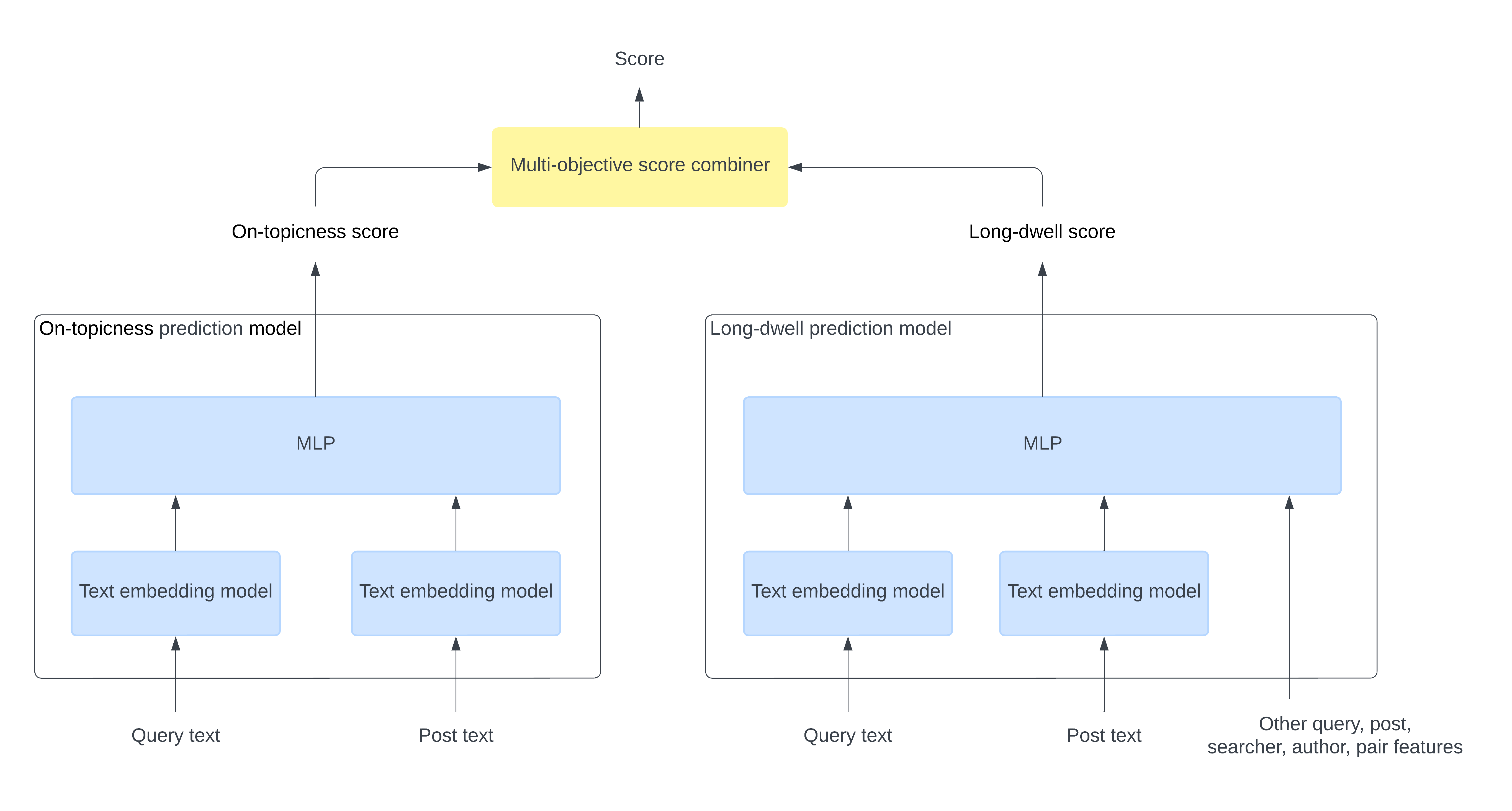}
	\caption[]{Architecture of the models used in L1 and L2 ranking stages.}
	\label{fig:ranking}
\end{figure*}

The models used in L1 and L2 have a similar architecture (as shown in Figure~\ref{fig:ranking}) and only differ in their size and the number of input features.

The architecture consists of two models, one for predicting the on-topicness score for each post and one for predicting the long-dwell score for each post. 

The on-topicness prediction model takes the following features as input:

\begin{itemize}
    \item Query text
    \item Post text
\end{itemize}

The query text and post text are passed to a text embedding model (multilingual-e5) which produces embeddings of the query text and post text. These text embeddings are concatenated and passed to an MLP which produces an on-topicness score.

The long-dwell prediction model take the following features as input:

\begin{itemize}
    \item Query text
    \item Post text
    \item (Query text, post text) pair features such as the BM25 match score
    \item Other query features, such as whether the query contains a job title
    \item Other post features, such as the popularity of the post
    \item Searcher features, such as whether the searcher has a job-seeking intent
    \item Author features, such as the popularity of the author
    \item (Searcher, author) pair features, such as whether the author and searcher are connected
\end{itemize}

Like in the on-topicness prediction model, the query text and post text are passed to a text embedding model (multilingual-e5) which produces embeddings of the query text and post text. These text embeddings are concatenated with the other features and passed to an MLP which produces a long-dwell score.

We again chose multilingual-e5 for embedding text in our ranking models as it also scores well on the MTEB leaderboard for the reranking task. Note that the usage of text embeddings in the ranking models enables semantic matching between the query and posts.

We train the on-topicness and long-dwell prediction models using (query, post, label) data collected from historical serving of posts for queries issued to the content search engine and their associated on-topicness/long-dwell labels.

The on-topicness score and the long-dwell score are then combined to produce a final score as follows:

\begin{equation}
	\text{score} = \alpha \text{ on-topicness score} + (1 - \alpha) \text{ long-dwell score}
\end{equation}

The parameter $\alpha$ here serves as a tuning knob to strike a desirable balance between the two objectives. We select $\alpha$ through online experimentation.

\section{Efficient serving}
To run the above content search engine online we make some other optimizations to ensure that the entire retrieval and ranking can be completed within a reasonable latency even at a high QPS. Some of these are:

\begin{enumerate}
    \item A limit on the number of posts to scan during the approximate nearest neighbor search in EBR. We experiment with several values and select the highest value allowed by our latency budget. 
    \item Precomputation of text embeddings of all the posts for usage in the ranking models. We precompute the text embeddings of all the posts (through offline and nearline pipelines) and push them to a key-value Venice \cite{venice} store. This not only reduces the latency of the ranking models, but also the amount of feature data that needs to be passed to the ranking models (as some post texts can be very long). 
\end{enumerate}

\section{Outcome}
Our new content search engine with semantic capability has made it possible to answer complex queries like "how to ask for a raise?", and improved on-topic rate and long-dwells by more than 10\%. We also observed a positive impact on LinkedIn's sitewide sessions, as members are more likely to engage with the platform when they get better search results.

\section{What's next?}
Although the on-topic rate metric does a reasonable job at capturing how a member would perceive the basic quality of posts, its simple definition does not capture all the expectations one would have for the quality of posts for the various types of queries. So, we are evolving the on-topic rate metric to a new metric that would better capture those expectations. Since optimizing this new metric requires a deep understanding of language, we plan to leverage an LLM in the ranking layer that jointly attends to the query and post text. We hope to share our learnings on this in the coming months.

\section{Acknowledgments}
We would like to thank Siddharth Singh, Justin Zhang, Ali Hooshmand, Sarang Metkar, Anand Kishore, Jiahao Xu, Andrew Tai, T.J. Hazen, Peng Yan, Jake Mannix, Vinay Krishnamurthy, Hongzhou Li, Hari Menon, Deepak Manoharan, Qing Li, Jin Sha, Shangjin Zhang, Ryan Kuk, Dianya Jiang, Viral Shah, Senthil Kumar, Jack Wang, Pratik Dixit, Sonali Garg, Ankur Agrawal, Lawrence Lin, Souvik Ray, Jitendra Agarwal, Manish Khanna, Dawn Woodard, Sean Henderson, Jeffrey Gochman, Dhruv Saksena, Jeffrey Wang, Birjodh Tiwana, Abhimanyu Lad, Alice Xiong, Rashmi Jain, Ginger Liu, Tim Jurka, Ya Xu, and many others for their contributions and support.

%
\bibliographystyle{ACM-Reference-Format}
\bibliography{biblio}


\begin{thebibliography}{00}


\ifx \showCODEN    \undefined \def \showCODEN     #1{\unskip}     \fi
\ifx \showDOI      \undefined \def \showDOI       #1{#1}\fi
\ifx \showISBNx    \undefined \def \showISBNx     #1{\unskip}     \fi
\ifx \showISBNxiii \undefined \def \showISBNxiii  #1{\unskip}     \fi
\ifx \showISSN     \undefined \def \showISSN      #1{\unskip}     \fi
\ifx \showLCCN     \undefined \def \showLCCN      #1{\unskip}     \fi
\ifx \shownote     \undefined \def \shownote      #1{#1}          \fi
\ifx \showarticletitle \undefined \def \showarticletitle #1{#1}   \fi
\ifx \showURL      \undefined \def \showURL       {\relax}        \fi
\providecommand\bibfield[2]{#2}
\providecommand\bibinfo[2]{#2}
\providecommand\natexlab[1]{#1}
\providecommand\showeprint[2][]{arXiv:#2}

\bibitem[\protect\citeauthoryear{??}{sam}{[n. d.]}]%
        {samza}
 \bibinfo{year}{[n. d.]}\natexlab{}.
\newblock \bibinfo{title}{Apache Samza}.
\newblock   (\bibinfo{year}{[n. d.]}).
\newblock
\newblock
\shownote{\url{https://samza.apache.org/}.}


\bibitem[\protect\citeauthoryear{??}{mte}{[n. d.]}]%
        {mteb}
 \bibinfo{year}{[n. d.]}\natexlab{}.
\newblock \bibinfo{title}{Massive Text Embedding Benchmark (MTEB) Leaderboard}.
\newblock   (\bibinfo{year}{[n. d.]}).
\newblock
\newblock
\shownote{\url{https://huggingface.co/spaces/mteb/leaderboard}.}


\bibitem[\protect\citeauthoryear{GV}{GV}{2022}]%
        {venice}
\bibfield{author}{\bibinfo{person}{Felix GV}.} \bibinfo{year}{2022}\natexlab{}.
\newblock \bibinfo{title}{Open Sourcing Venice - LinkedIn's Derived Data
  Platform}.
\newblock   (\bibinfo{year}{2022}).
\newblock
\newblock
\shownote{\url{https://www.linkedin.com/blog/engineering/open-source/open-sourcing-venice-linkedin-s-derived-data-platform}.}


\bibitem[\protect\citeauthoryear{Wang, Yang, Huang, Yang, Majumder, and
  Wei}{Wang et~al\mbox{.}}{2024}]%
        {e5}
\bibfield{author}{\bibinfo{person}{Liang Wang}, \bibinfo{person}{Nan Yang},
  \bibinfo{person}{Xiaolong Huang}, \bibinfo{person}{Linjun Yang},
  \bibinfo{person}{Rangan Majumder}, {and} \bibinfo{person}{Furu Wei}.}
  \bibinfo{year}{2024}\natexlab{}.
\newblock \bibinfo{title}{Multilingual E5 Text Embeddings: A Technical Report}.
\newblock   (\bibinfo{year}{2024}).
\newblock
\showeprint[arxiv]{cs.CL/2402.05672}
\showURL{%
\url{https://arxiv.org/abs/2402.05672}}


\end{thebibliography}

\end{document}